\documentclass[twocolumn,twoside,slac_two]{revtex4}
\usepackage{graphicx}
\usepackage{fancyhdr}
\usepackage{hyperref}
\hypersetup{
    colorlinks=true,
    urlcolor=magenta,
}
\begin{document}

\title{The Extreme Energy Events HECR array: status and perspectives}

%

\author{I. Gnesi$^{a,f}$, M. Abbrescia$^{a,b}$, C. Avanzini$^{a,c}$, L. Baldini$^{a,c}$, R. Baldini Ferroli$^{a,d}$, G. Batignani$^{a,c}$, G. Bencivenni$^{d}$, E. Bossini$^{a,e}$, A. Chiavassa$^{f}$, C. Cicalo$^{a,g}$, L. Cifarelli$^{a,h}$, F. Coccetti$^{a}$, E. Coccia$^{i}$, A. Corvaglia$^{a,j}$, D. De Gruttola$^{a,k}$, S. De Pasquale$^{a,k}$, A. Di Giovanni$^{l}$, M. D'Incecco$^{l}$, M. Dreucci$^{d}$, F.L. Fabbri$^{d}$, E. Fattibene$^{m}$, A. Ferraro$^{m}$, V. Frolov$^{p}$, P. Galeotti$^{a,f}$, M. Garbini$^{a,h}$, G. Gemme$^{q}$, S. Grazzi$^{a,q}$, C. Gustavino$^{l}$, D. Hatzifotiadu$^{a,h,o}$, F. Liciulli$^{a,v}$, P. La Rocca$^{a,r}$, A. Maggiora$^{f}$, O. Maragoto Rodriguez$^{t}$, G. Maron$^{m}$, B. Martelli$^{m}$, M.N. Mazziotta$^{s}$, S. Miozzi$^{a,d,i}$, R. Nania$^{a,h}$, F. Noferini$^{a,m}$, F. Nozzoli$^{i,t,n}$, M. Panareo$^{a,j}$, M.P. Panetta$^{a,j}$, R. Paoletti$^{a,e}$, W. Park$^{n}$, L. Perasso$^{a,q}$, F. Pilo$^{a,c}$, G. Piragino$^{a,f}$, F. Riggi$^{a,r}$, G.C. Righini$^{a}$, M. Rizzi$^{v}$, G. Sartorelli$^{a,h}$,
E. Scapparone$^{h}$, M. Schioppa$^{a,u}$, A. Scribano$^{a,c}$, M. Selvi$^{h}$, S. Serci$^{f}$, E. Siddi$^{g}$, S. Squarcia$^{q}$, L. Stori$^{a,g}$, M. Taiuti$^{q}$, G. Terreni$^{c}$, O.B.Visnyei$^{n}$, M.C. Vistoli$^{m}$, L. Votano$^{l}$, M.C.S. Williams$^{e,o}$, S. Zani$^{m}$, A. Zichichi$^{a,h,o}$, R. Zuyeuski$^{a,o}$}
\affiliation{a - Museo Storico della Fisica e Centro Studi e Ricerche E. Fermi, Roma, Italy\\
b - INFN and Dipartimento di Fisica Universit\`a di Bari, Italy\\
c - INFN and Dipartimento di Fisica Universit\`a di Pisa, Italy\\
d - INFN Laboratori Nazionali di Frascati (RM), Italy\\
e - INFN Gruppo Collegato di Siena and Dip. di Fisica Univ. di Siena, Italy\\
f - INFN and Dipartimento di Fisica Universit\`a di Torino, Italy\\
g - INFN and Dipartimento di Fisica Universit\`a di Cagliari, Italy\\
h - INFN and Dipartimento di Fisica Universit\`a di Bologna, Italy\\
i - INFN and Dipartimento di Fisica Universit\`a di Roma Tor Vergata, Italy\\
j - INFN and Dipartimento di Matematica e Fisica Universit\`a del Salento - Lecce, Italy\\
k - INFN and Dipartimento di Fisica Universit\`a di Salerno, Italy\\
l - INFN Laboratori Nazionali del Gran Sasso - Assergi (AQ), Italy\\
m - INFN-CNAF Bologna, Italy\\
n - ICSC World Laboratory Geneva, Switzerland\\
o - CERN, Geneva, Switzerland\\
p - JINR Joint Institute for Nuclear Research, Dubna, Russia\\
q - INFN and Dipartimento di Fisica Universit\`a di Genova, Italy\\
r - INFN and Dipartimento di Fisica e Astronomia Universit\`a di Catania, Italy\\
s - INFN sezione di Bari, Italy\\
t - ASI Science Data Center Roma, Italy\\
u - Dipartimento di Fisica Universit\`a della Calabria - Cosenza, Italy\\
v - INFN Sezione di Bari, Italy\\
}

\begin{abstract}
The Extreme Energy Events Project is a synchronous sparse array of 52 tracking detectors for studying High Energy Cosmic Rays (HECR) and Cosmic Rays-related phenomena. The observatory is also meant to address Long Distance Correlation (LDC) phenomena: the network is deployed over a broad area covering 10 degrees in latitude and 11 in longitude.
An overview of a set of preliminary results is given, extending from the study of local muon flux dependance on solar activity to the investigation of the upward-going component of muon flux traversing the EEE stations; from the search for anisotropies at the sub-TeV scale to the hints for observations of km-scale Extensive Air Shower (EAS). 
\end{abstract}

\maketitle

\thispagestyle{fancy}


\section{Introduction}
The Extreme Energy Events \citep{CF, Zich, EEEinit} observatory is a regularly growing net of high time resolution tracking detectors. The stations are spread over more than 3$\times$10$^5$ km$^2$. Each site hosts
a muon tracking telescope made of three MRPCs, very similar, conceptually, to the chambers
developed for the Time-Of-Flight system of the ALICE experiment at LHC \citep{AliceTof, Akin}.
The net is organized both in clusters and single telescope sites, in order to address Cosmic Ray (CR) phenomenon over a broad energy range. 

EEE has also a strong outreach impact. Most of the detectors are installed in high schools where students and
teachers actively participate to the data taking activities, taking care of the operation and manteinance of their telescopes. In the last three years more than 100 teachers and several thousands of students have been involved
in the Project. Researchers coordinate and supervise activities, providing
support during the construction, installation and use of the detectors. Students
and teachers are introduced through seminars, lectures and masterclasses to the scientific research community, with the opportunity of understanding how a real experiment works, from the infrastructure development to the data acquisition, analysis and publications of scientific results.

\begin{figure}[!ht]
\centering
\includegraphics[width=65mm]{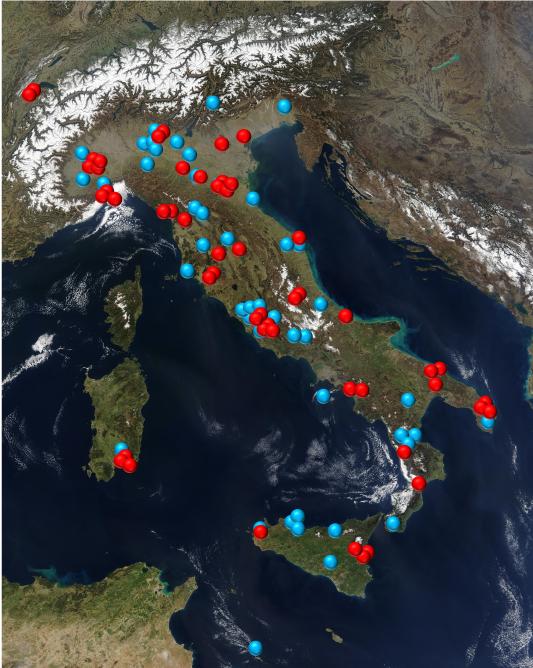}
\caption{EEE sites. Red points: sites where a telescopes is installed. Blue points: Schools in waiting list for the telescopes} \label{figMap}
\end{figure}

In Figure \ref{figMap} is showed an up-to-date map of the EEE detection sites
(in red). Schools involved in the Project without hosting the detector are plotted in blue. These sites are possible future telescope installations. The result is a sparse array where each detection site is made of few detectors at 30 m - 2 km distance from the others. Currently
the largest distance between the EEE telescopes is about 1200 km (between CERN laboratories and
Catania).

\section{The EEE Station}
The EEE telescope allows the detection of EAS muons with high efficiency and good angular resolution. Each telescope is composed by three MRPC chambers of $160\times80$ $cm^{2}$ size, a simplified and larger version of the detectors developed for the time-of-flight system of the Alice experiment. Each MRPC consists of six gas gaps obtained interleaving two glasses, coated with resistive paint and acting as electrodes, with a stack of five floating glasses. Commercial nylon fishing line (300 $\mu$m thick) is used as spacer between glasses. The chambers are filled with a mixture of 98\% of C2H2F4 and 2\% of SF6 and are operated in avalanche mode.

\begin{figure}[!ht]
\centering
\includegraphics[width=80mm]{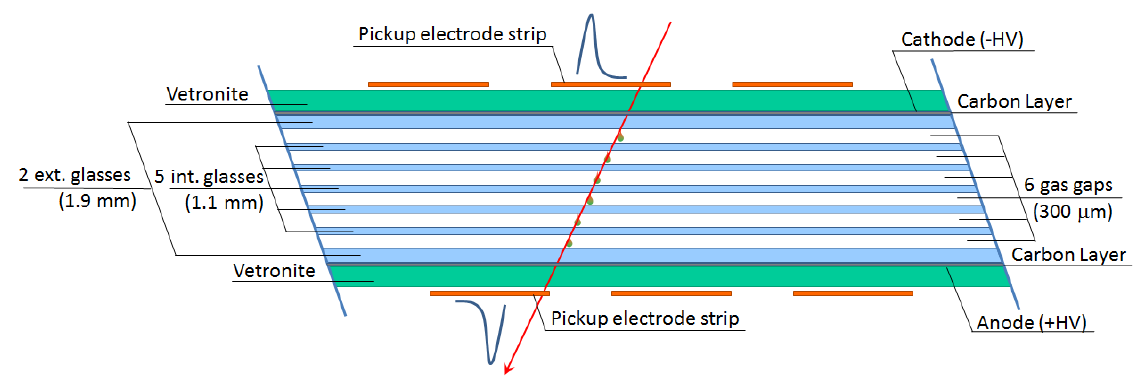}
\caption{Cross section of a MRPC plane of the EEE telescope.} \label{figChamber}
\end{figure}

The MRPC pick-up electrodes are segmented into 24 strips read out at the two ends. This strip configuration is used to provide two-dimensional information when a particle crosses the chamber: one coordinate is given by the fired strip, while the second is obtained by the time difference of the arrival signals at the opposite ends in each strip (see Figure \ref{figChamber}). The strips are read out by front-end (FEA) cards, which incorporate an ultra-fast and low-power amplifier/discriminator ASIC specifically designed for MRPC operation \cite{NINOChip}. From the front-end cards a total of 144 LVD digital signals are fed into two commercial multi-hit TDCs which provide time measurements with a 100 ps precision in a time window of 500 ns; this solution provides a value of about 70 ps for the time resolution and 0.7 cm for the spatial resolution along the readout strips direction, while in the transverse direction the measured spatial resolution is about 0.25 in units of strips, corresponding to 0.63 cm \cite{EEESpaRes}.
The telescope data acquisition system is VME based, controlled by a LabView interface running on a PC connected to the VME crate via a USB-VME bridge. GPS interfacing and trigger logic complete the set of boards hosted in the VME crate. The 0 level trigger is the coincidence among the 6 all-strip OR signals given by front-end cards, one per each side of the 3 MRPCs. The trigger logic brings to a suppression of the spurious events below 1 Hz, even if the MRPCs dark rate is generally well above 10 kHz. GPS allows the synchronization of the different EEE stations with a precision better than 30 ns.
\section{Data Taking and Monitoring}
The EEE Observatory data taking is organized in coordinated runs, during which the telescopes are regularly monitored for achieving the highest possible overall duty cicle. The simultaneous data takings started with a three-week PILOT run at the end of 2014. Then the RUN1 was successfully performed in 2015, with
about 35 telescopes for a two-month period (March-April 2015) and an overall number of collected events around 5 billion; the second combined data taking, RUN2, started in October 2015 and lasted until end of May 2016, with 40 telescopes involved and 15 billion of events collected. The next run, RUN3, is starting in October 2016 and will last until June 2017 allowing the collection of a statistics a factor 2 higher then the previous RUNs.\\The EEE Project data reconstruction and quality monitoring is implemented at the CNAF \cite{CNAF} cloud facility, where a OpenStack-based environment allows to allocate resources for both collecting and storing data from the telescopes and for running the track reconstruction algorithm and the subsequent Data Quality Monitor system. Quality plots are made available to the Collaboration and to the schools participating to the experiment. An automated monitor produces a daily report describing the current status of the telescopes. Each telescope is also regulary monitored by groups of students and teachers trained to the purpose by researchers of the EEE Collaboration; the telescope status is then saved to an online e-logbook. Information from both the DQM and e-logbook are used for data selection during analysis and for telescope manteinance.
\section{Results}
The main topics addressed by the EEE Project are related to EAS, i.e. short and far distance
coincidences detection and also low energy Cosmic Ray - related phenomena. 
\subsection{EAS}
\begin{figure}[!ht]
\centering
\includegraphics[width=80mm]{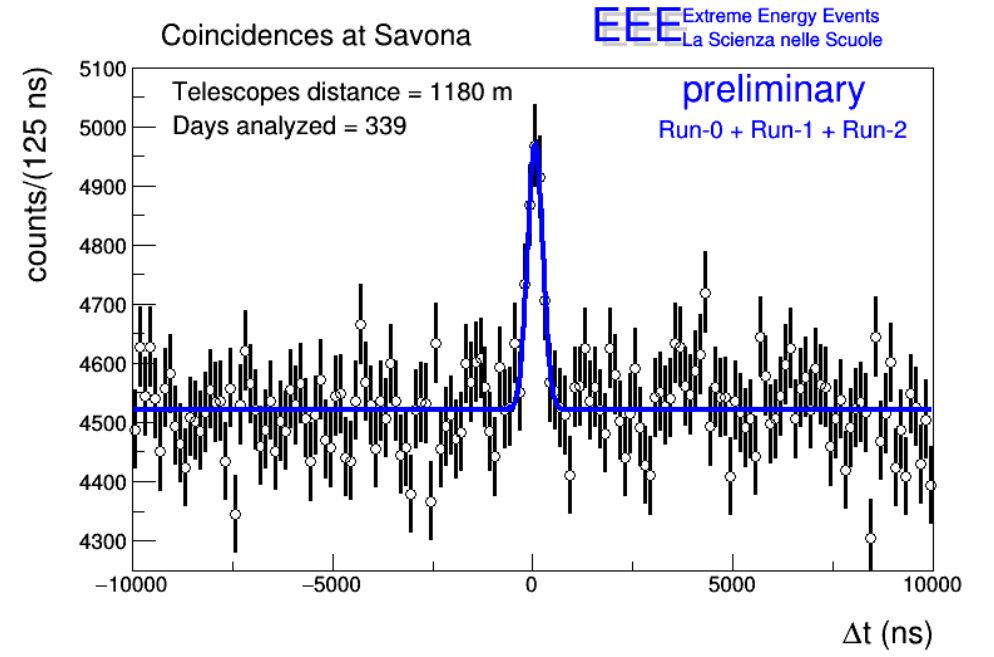}
\caption{Time difference distribution between two tracks observed by different EEE telescopes placed at a relative distance of 1.2 km in Savona.} \label{figSAVO}
\end{figure}

The topology of the EEE network is thought in order to open the possibility of measuring time-correlated events at distances never addressed before. Stations are then organized in 2-4 stations local clusters, where the main scope is the measurement of knee and sub-GZK EAS, depending on the typical size of the clusters. Clusters are then installed apart, providing a range of inter-distances which span from few tenth of km to more than 1000 km for the farthest stations.
Telescopes placed in the same city can detect individual extensive air showers, whereas telescopes located hundreds of kilometers apart can, in principle, detect the coincidence between two different correlated air showers. The last is expected to be an extremely rare phenomenon, since no experimental evidence of such events has been observed so far. On the other side the extremely low rate expected is foreseen by models which involve super-massive DM decays \cite{DM1,DM2} with very long life-time, topological defects, decays of monopoles bound states \cite{Monopoles} and evaporation of primordial black holes \cite{BH}. These scenarios open a wide set of possible relative distances for correlated EAS and their flux.\\
A huge statistics is therefore necessary in order to reduce the statistical
uncertainties and increase the signal-background ratio.\\
On the other side, the detection of single EASs has been performed using telescopes clusters placed in the same city. Figure \ref{figSAVO} shows, as an example, the time-difference spectrum obtained
between two telescopes in Savona placed at a relative distance of about 1.2 km. The significance is $\frac{S}{\sqrt{S+B}}\sim$9.7. Thanks to tracking and precise timing information the time difference in figure \ref{figSAVO} has been corrected, event by event, for the time delay between the two telescopes due to the different path travelled by particles of the EAS disk.

The increasing statistics is opening up the possibility for new studies ({\it e.g.} the
search for long-range correlations between EASs) while improving the significance of single EAS observation.
\subsection{Anisotropies}
Galactic cosmic rays are expected to be nearly isotropic, due to their interaction with the galactic magnetic field over the long path to the Earth. Observation of cosmic rays at energies smaller than 10$^{15}$ eV is a useful tool to inspect the magnetic field structure in the interstellar medium of our Solar System. Small anisotropies may be induced by large scale as well as local magnetic field features. At 1 TeV, sizeable effects from the eliosphere may be expected on the cosmic ray distributions.
The so-called Compton-Getting effect, introduced in 1935 \cite{CG}, predicts a dipole component in the cosmic ray anisotropy, due to the relative motion of the observer with the isotropic cosmic ray plasma rest frame: {\it i.e.} the anisotropy due to Earth motion around the Sun has been observed by several experiments to be of the order of $\sim$10$^{-4}$.

A first analysis with a dataset collected by the EEE network has been performed, providing a skymap in equatorial coordinates, for the dominating sub-TeV extensive air showers. More than 100 M events  were fully reconstructed after track quality selection. Raw data were corrected for the time exposure of each
telescope and its geometrical acceptance by the scrambling method, and corrected maps were extracted and plotted in the Right Ascension vs Declination coordinates. Corrected data maps
are compatible with isotropic distributions at the level of 5$\times$10$^{-3}$ - 10$^{-2}$, being compatible with statistical fluctuations associated to the number of analysed events.
The available statistics is now 2 order of magnitudes larger than the one used for obtaining these results, allowing for inhomogeneities down to 10$^{-3}$. Additional analyses are being performed to search for possible deviations from an isotropic pattern at a deeper confidence level.
\subsection{Galactic Cosmic Ray Decreases (GCRD)}
\begin{figure}[!ht]
\centering
\includegraphics[width=80mm]{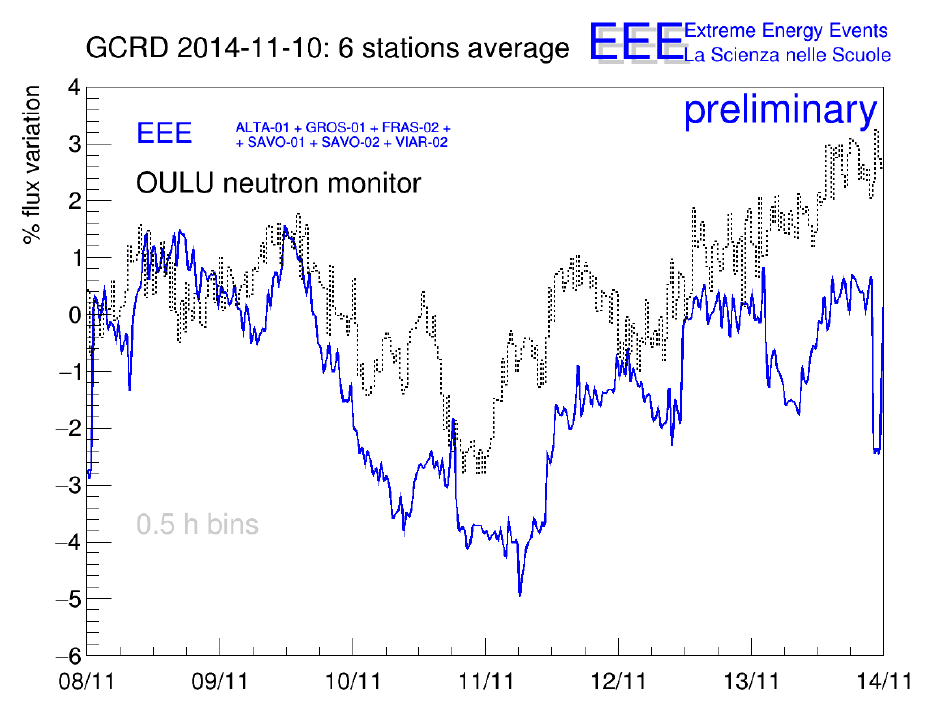}
\caption{Blue: GCRD observed by 6 EEE teelscopes. Black: OULU Neutron Monitor data.} \label{figGCRD}
\end{figure}

The monitoring of galactic cosmic ray flux decreases is of interest for understanding phenomena occurring on the solar heliosphere, as well as on other observable stars. As it is known, they are related to the emission of mass from star corona and often to solar flares, even
if such relation is not completely understood, especially the interplay with interplanetary structures
\cite{GCRD1,GCRD2}. The effect on the solar wind directly affects the measured galactic cosmic ray flux on Earth, giving typical flux fluctuations of a few percent on a few days basis.\\
The long term survey of cosmic ray flux fluctuations has been historically performed by neutron
monitors (NM), showing usually higher angular acceptance and lower energy threshold than EEE muon tracking detectors. On the other side the overall acceptance of each EEE station, the timing resolution better than 30 ns and the tracking capabilities allow for the study of the muon component of the GCRDs, opening to comparison with NM and to a deeper understanding of such phenomena.\\
Four variations have been already observed by the EEE telescopes in 2011, 2012, 2014 and 2015, by correcting the absolute flux for the known sources of systematic uncertainties, mainly due to pressure fluctuations above the telescopes. The results show that the EEE array has the capability of becoming a stable survey for GCRDs over a broad latitude and longitude range.
In Figure \ref{figGCRD} a GCRD occurred in November 2014 and observed by 6 EEE stations is shown in comparison with the OULU NM.
\section{Conclusions}
The EEE network represents a unique case where a high performance and widespread CR observatory plays a fundamental role in science outreach, involving teacher a students in real research activities. The excellent performances of the stations are confirmed by first results on EAS observations, low energy CR anisotropies and solar activity survey. The growing size of the network and the implementation of new analysis techniques is being pursued for addressing complex open physics questions as the search for long distance correlations.
\bigskip 

\end{document}